\documentclass[suppldata]{interact}
\usepackage{epstopdf}
\usepackage[caption=false]{subfig}

\usepackage[numbers,sort&compress,merge]{natbib}
\bibpunct[, ]{[}{]}{,}{n}{,}{,}
\renewcommand\bibfont{\fontsize{10}{12}\selectfont}

\usepackage{amsmath,amsfonts,amsthm}
\usepackage[dvipsnames]{xcolor}
\usepackage{graphicx}
\usepackage{multirow,chemformula}
\usepackage[version=4]{mhchem}
\usepackage[colorlinks=true,breaklinks=true,linkcolor=magenta,citecolor=magenta,urlcolor=magenta]{hyperref}
\usepackage{comment}
\usepackage{array}
\usepackage{booktabs}
\usepackage{tablefootnote}

\theoremstyle{plain}

\theoremstyle{definition}

\theoremstyle{remark}

\newcommand{\device}[1]{$\mathsf{ibm\_#1}$}
\newcommand{\library}[1]{$\mathsf{#1}$}

\newcommand{\bgreek}[1]{{\boldsymbol{#1}}}
\newcommand{\vett}[1]{{\bf{#1}}}
\newcommand{\parm}{\bgreek{\theta}}

\newcommand{\bs}{\vett{x}}
\newcommand{\crt}[1]{\hat{c}_{#1}^\dagger}
\newcommand{\dst}[1]{\hat{c}_{#1}^{\phantom{\dagger}}}
\newcommand{\oper}[1]{\hat{#1}}

\begin{document}

\title{Quantum computation of $\pi \to \pi^*$ and $n \to \pi^*$ excited states of aromatic heterocycles}

\author{
\name{Maria A. Castellanos\textsuperscript{a*} \thanks{\textsuperscript{*}mariacm@mit.edu},
Mario Motta\textsuperscript{b} and Julia E. Rice\textsuperscript{b**} \thanks{\textsuperscript{**}jrice@us.ibm.com}}
\affil{\textsuperscript{a} Department of Chemistry, Massachusetts Institute of Technology, Cambridge, MA 02139, USA; \textsuperscript{b} IBM Quantum, IBM Research - Almaden, 650 Harry Road, San Jose, CA 95120, USA}
}

\maketitle

\begin{abstract}

The computation of excited electronic states is an important application for quantum computers.
In this work, we simulate the excited state spectra of four
aromatic heterocycles on IBM superconducting quantum computers, focusing on active spaces of $\pi \to \pi^*$ and $n \to \pi^*$ excitations.
We approximate the ground state with the entanglement forging method, a qubit reduction technique that maps a spatial orbital to a single qubit, rather than two qubits. We then determine excited states using the quantum subspace expansion method. We showcase these algorithms on quantum hardware using up to 8 qubits and employing readout and gate error mitigation techniques.
Our results demonstrate a successful application of quantum computing in the simulation of active-space electronic wavefunctions of substituted aromatic heterocycles, and outline challenges to be overcome in elucidating the optical properties of organic molecules with hybrid quantum-classical algorithms.
\end{abstract}

\begin{keywords}
Quantum computation; electronic excited states; aromatic heterocycles
\end{keywords}

\section{Introduction}

The investigation of molecular excited states has significant implications across various fields, including photosynthesis~\cite{turro2009principles,Scholes2011lessons}, photovoltaics~\cite{Bredas2016photovoltaic} and catalysis~\cite{welin2017photosensitized,nilsson2005electronic}.
Aromatic rings are common motifs in molecules involved in biochemical processes, such as photosynthetic pigments, nucleotides, drugs, and certain amino acids. 
The typical UV/Vis absorption spectra (200-800 nm) of aromatic molecules are characterized by intense bands attributed to $\pi\rightarrow \pi^*$ and $n\rightarrow \pi^*$ valence transitions~\cite{harris1989symmetry}, although Rydberg transitions involving excitations to diffuse orbitals may also be observed in low-energy absorption spectra~\cite{Serrano-Andres1993,Baranov1986}. Valence excitations often drive biologically-relevant photochemical reactions involving aromatic molecules.

Electronic structure methods for excited states typically fall in two categories~\cite{Dreuw2005single-ref}: those derived from the Hartree-Fock (HF) formalism, such as configuration interaction (CI)~\cite{Foresman1992toward,HeadGordon1994doublescorr}, complete active-space second-order perturbation theory (CASPT2)~\cite{andersson1990second, andersson1992second}, and equation-of-motion coupled-cluster theory (EOM-CC)~\cite{Krylov2008eom-ccsd,Sekino1984linear,christiansen1995response}, and those based on time-dependent density functional theory (TD-DFT)~\cite{Dreuw2005single-ref}. 
While CC-based methods can offer an accurate description of the excited-state electronic structure, their high computational cost often limits their applicability to small molecules. More cost-efficient approaches, such as TD-DFT and configuration interaction singles (CIS) and singles and perturbed doubles (CIS(D))~\cite{head1994doubles}, can be used for larger systems.

Recent years have witnessed the development of quantum computing algorithms \cite{georgescu2014quantum,cao2019quantum,bauer2020quantum,mcardle2020quantum,motta2021emerging} for studying ground and excited states of electronic systems. These include variational methods \cite{kandala2017hardware,higgott2019variational,gao2021applications} and subspace methods like the quantum subspace expansion (QSE) \cite{mcclean2017subspace,colless2018computation,smart2021quantum} and the quantum equation-of-motion \cite{ollitrault2020quantum}, which build upon a ground-state approximation to calculate excited-state energies and properties \cite{colless2018computation,gao2021applications,huang2022variational,motta2023quantum}.
While promising and rapidly progressing, these approaches have a number of challenges:
(i) like classical algorithms, quantum computing algorithms are based on approximation schemes.
(ii) due to the intrinsically probabilistic nature of quantum measurements, the output of quantum computing algorithms is accompanied by statistical uncertainties (a phenomenon called ``shot noise'' in quantum computing literature). 
(iii) the simulation of quantum computing algorithms on near-term devices is limited by the number of qubits that can be reliably prepared, manipulated, and measured.
In light of these challenges, it is important to quantify the impact of approximations, shot noise, and devices on results produced via quantum computing algorithms, particularly when studying the electronic structure of molecules.

With these challenges in mind, we investigated the low-lying valence excited states of four common aromatic molecules, namely furan, pyrrole, pyridine, and pyrimidine. We constructed an active space of $\pi\rightarrow \pi^*$ and $n\rightarrow \pi^*$ valence transitions in which
we approximated the ground-state wavefunction with 
a variational quantum computing algorithm based on a qubit reduction
technique, called entanglement forging (EF) \cite{eddins2022doubling,motta2023quantum}.
We then approximated the excited states 
with a QSE based on single and double electronic excitations applied to the ground-state wavefunction yielded by EF~\cite{mcclean2017subspace,colless2018computation}.

The goal of this work is to assess the accuracy of quantum computing algorithms for characterizing molecular ground and excited states within an active space (exemplified by the combination of EF and QSE) using substituted aromatic rings as a relevant example.
Success in this endeavor is an important step toward the calculation of high-accuracy excitation energies using quantum algorithms and quantum computing hardware.
It also paves the way for more realistic applications with relevance in organic chemistry and biological applications.

The remainder of this paper is organized as follows. In the Methods section, we describe the EF and QSE algorithms for ground- and excited-state simulations and provide details on the excited-state calculations.
In the Results section, we illustrate calculations conducted on classical simulators and quantum hardware, and compare them to state-of-the-art electronic structure simulations and experimental results.

In the Discussion section, we analyze the results in terms of the three challenges outlined above.
Finally, in the Conclusions section, we summarize the results and identify areas of future research.

\section{Methods}

\subsection{Entanglement forging (EF) for ground-state simulations}

In this work, we performed ground-state calculations as a starting point for the subsequent determination of excited states. For each active space (see ``Calculation Details'' for their construction), we approximated the ground-state wavefunction using the entanglement forging (EF) technique~\cite{eddins2022doubling,motta2023quantum}.
One of the main challenges for near-term quantum algorithms,
and the motivation behind EF, lies in the resources required for mapping molecular operators and wavefunctions onto qubits. In general, simulating electrons in $N$ molecular orbitals (MOs) requires $2N$ qubits (two for each MO, since each MO can be empty, occupied by an $\alpha$ electron, occupied by a $\beta$ electron, or doubly occupied, and these 4 states are described by 2 qubits).
Consequently, simulating systems of interest often exceeds the budget of contemporary hardware.
Within EF, a qubit represents a spatial-orbital rather than a spin-orbital, and thus the number of qubits required for a simulation is reduced by half (from $2N$ to $N$).
This qubit reduction is achieved by representing the target wavefunction (here, a ground-state wavefunction) by a Schmidt decomposition,
\begin{equation}
\label{eq:ef_target}
| \Psi_{\parm} \rangle = \sum_k \lambda_k \, 
\oper{U}(\parm) | \bs_k \rangle \otimes \oper{U}(\parm) | \bs_k \rangle    
\end{equation}
where the tensor product separates a $2N$-qubit register in two $N$-qubit registers, respectively for $\alpha$ and $\beta$ spin-orbitals, $\oper{U}(\parm)$ is a parameterized unitary, $\lambda_k$ a set of Schmidt coefficients, and $| \bs_k \rangle$ are 
qubit computational basis states represented as binary strings.
We write the active-space Hamiltonian as a linear combination of tensor products,
\begin{equation}
\oper{H} = \sum_\mu \oper{A}_\mu \otimes \oper{B}_\mu 
\quad,
\end{equation}
where $\oper{A}_\mu$ and $\oper{B}_\mu$ act on $\alpha$ and $\beta$ spin-orbitals respectively. We then define its expectation value,
\begin{equation}
\label{eq:ef_1}
\langle \Psi_{\parm} | \oper{H} | \Psi_{\parm} \rangle 
= 
\sum_{kl , \mu} \lambda_k \lambda_l \, A_{kl\mu} B_{kl\mu}
\;,
\end{equation}
in terms of matrix elements of the form 
\begin{equation}
A_{kl\mu}
=
\langle \bs_k | \oper{U}^\dagger (\parm) \oper{A}_\mu \oper{U}(\parm) | \bs_l \rangle
\;,
\end{equation}
one for each term of the Hamiltonian (labeled by the index $\mu$) and pair of binary strings (labeled by indices $kl$). Evaluating matrix elements where the bra and the ket states differ is an expensive operation on a near-term quantum computer because, in general, it requires additional qubits (ancillae) and controlled operations~\cite{somma2002simulating}.
When the bra and ket states are computational basis states, the matrix element $A_{kl\mu}$ can be written as a linear combination of expectation values of $\hat{A}$ 
over superpositions of the states $| \bs_k \rangle$ and $| \bs_l \rangle$ with coefficients $i^p$, where $p=0,1,2,3$. More specifically,
\begin{equation}
\label{eq:ef_2}
\begin{split}
A_{kl\mu} 
&= \sum_{p=0}^3 \frac{(-i)^p}{4}
\langle \phi^p_{kl} | \oper{A}_\mu | \phi^p_{kl} \rangle
\,, \\
| \phi^p_{kl} \rangle 
&= \frac{ |{\bf{x}}_k \rangle + i^p | {\bf{x}}_l \rangle }{ \sqrt{2} }
\,.
\end{split}
\end{equation}
Within EF, one prepares the states $| \phi^p_{kl} \rangle$ on a quantum processor, and then
measures the matrix elements $A_{kl\mu}$, $B_{kl\mu}$ and the expectation value 
in Eq.~\eqref{eq:ef_1}.

A description of the specific states ${\bf{x}}_k$, ansatz $\oper{U}(\parm)$, and quantum circuits studied in this work is presented in the ``Calculation Details'' section.

\subsection{Quantum subspace expansion (QSE) for excited-state simulations}

To model excited states, we used a quantum subspace expansion (QSE) built upon the EF ground state~\cite{mcclean2017subspace,colless2018computation,motta2023quantum}.
Within the QSE, the time-independent Schr\"{o}dinger equation for ground and excited states is 
projected into a subspace of the many-electron Hilbert space, spanned by a basis of wavefunctions of the form $| e_\mu \rangle = \hat{E}_\mu | \Psi \rangle$,
\begin{equation}
\label{eq:qse_equation}
\sum_\nu H_{\mu\nu} \, c_{\nu A} = E_A \sum_\nu M_{\mu\nu} \, c_{\nu A} \;,
\end{equation}
where $H_{\mu\nu} = \langle e_\mu | \hat{H} | e_\nu \rangle$ and $M_{\mu\nu} = \langle e_\mu | e_\nu \rangle$ are the Hamiltonian and overlap matrix elements, respectively.
We remark that QSE does not give any specific prescription for the choice of basis vectors $| e_\mu \rangle$. These should strike a balance between computational cost, the accuracy of the resulting eigenpairs $\{ E_A,\sum_\mu c_{\mu A} |e_\mu \rangle \}_A$, and numerical stability determined by the condition number of the overlap matrix $M$.

In this work, we focused on single and double electronic excitations within the same set of orbitals used to describe the ground state, 
\begin{equation}
\label{eq:qse_basis}
\begin{split}
|e_0 \rangle &= | \Psi \rangle \; , \\
|e_{ai} \rangle &= \crt{a\sigma} \dst{i\sigma} | \Psi \rangle \; , \\
|e_{aibj} \rangle &= \crt{a\sigma} \crt{b\tau} \dst{j\tau} \dst{i\sigma} | \Psi \rangle \; , \\
\end{split}
\end{equation}
where $\dst{p\eta}$ ($\crt{p\eta}$) destroys (populates) an electron with spin $\eta \in \{\alpha,\beta\}$ at the spatial-orbital $p$.
We clarify that the choice in Eq.~\eqref{eq:qse_basis} leads to a MR-CISD method (multireference configuration interaction with singles and doubles). It is known that CI methods while variational, are neither size-extensive nor size-intensive. Therefore, deviations between exact and computed ground- and excited-state energies are expected for systems with more than two electrons, to an extent that increases with the number of active-space electrons and orbitals.
On the other hand, use of singles and doubles leads to well-conditioned overlap matrices (e.g. when $\Psi$ is the Hartree-Fock state, the basis vectors $|e_\mu \rangle$ are orthonormal) and only results in additional measurements on the EF wavefunction, albeit with a severe scaling with active-space size \cite{tammaro2022n,motta2023quantum}. This makes the choice of basis, Eq.~\eqref{eq:qse_basis}, acceptable for small active-space simulations on near-term quantum hardware.

It is very important to classify the resulting excited states based on total spin and irreducible representation (irrep). To characterize these states we: 
(i) Constructed a matrix $S_{\mu\nu} = \langle e_\mu | \hat{S}^2 | e_\nu \rangle$ for the total spin operator.
(ii) Extracted a block of the matrices $H_{\mu\nu}$, $S_{\mu\nu}$, and $M_{\mu\nu}$ for each irrep (e.g. $A_1$, $A_2$, $B_1$, and $B_2$) within the $C_{2v}$ symmetry group.
(iii) Within each irrep (corresponding to a suitable block of the matrices $H$, $S$, and $M$) we solved a generalized eigenvalue equation (GEEV) for the total spin operator to identify singlet/triplet eigenspaces, and then solved a GEEV for the Hamiltonian operator in each eigenspace. 
This series of operations allowed us to label ground and excited states based on their total spin and irrep.

\subsection{Calculation Details}

\begin{figure*}[t!]
\centering
\includegraphics[width=\textwidth]{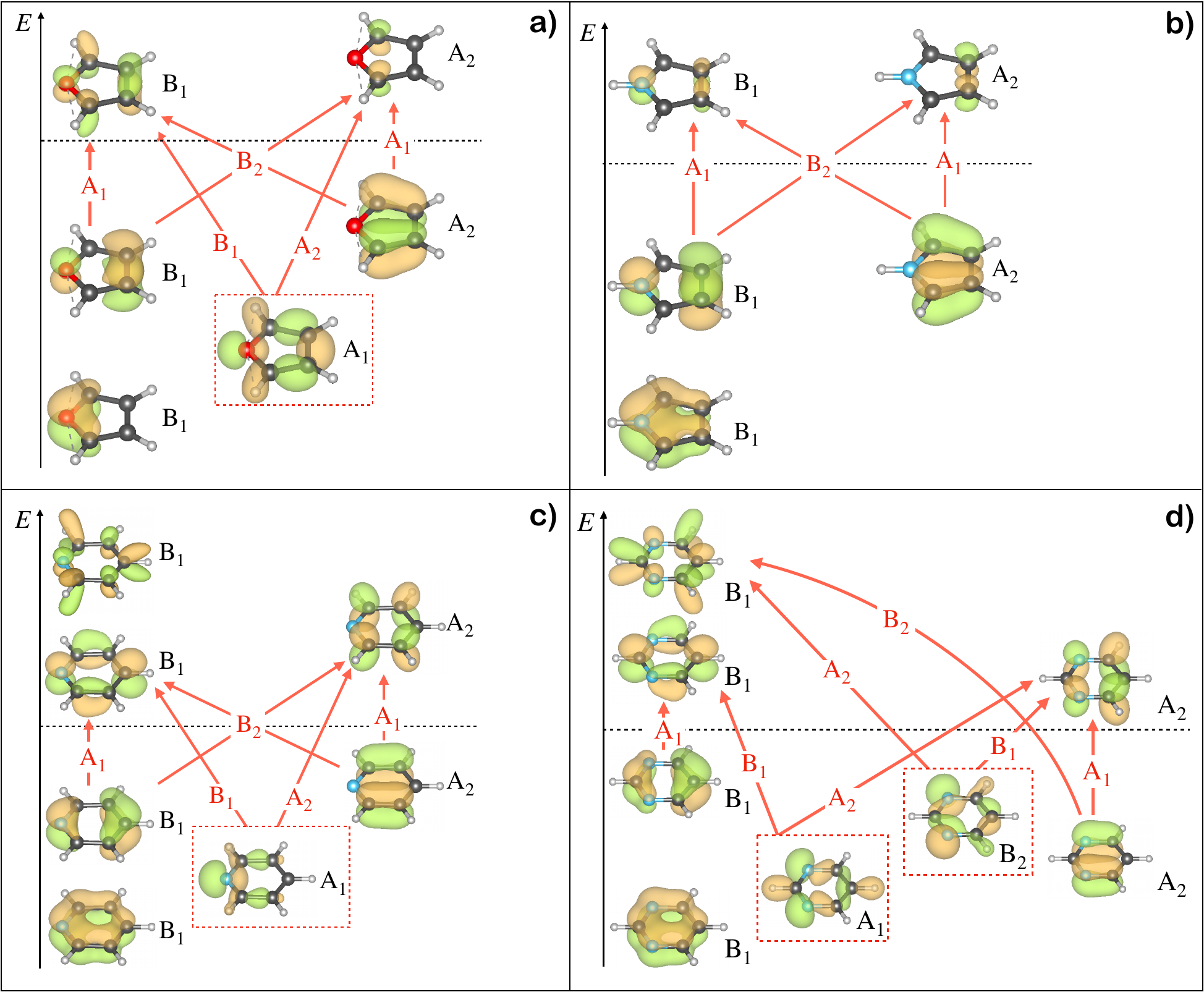}
\caption{Active-space MOs for furan (a), pyrrole (b), pyridine (c) and pyrimidine (d). The irreducible representation (irrep) of each MO is indicated in black, while one-electron excitations are indicated in red. The dashed black line separates occupied and virtual orbitals. The lone-pair orbital is enclosed in a dashed red rectangle. To avoid clutter, only excitations from the highest-lying occupied $\pi$ and lone-pair MOs to the lowest-lying $\pi^*$ MOs are shown.}
\label{fig:active_spaces}
\end{figure*}
\begin{table}[h!]
\centering
\begin{tabular}{cccccc}
\hline\hline
species & $\pi$ space & $n+\pi$ space \\
\hline
furan $(\ce{C4H4O})$       & (6e,5o) & (8e,6o)  \\
pyrrole $(\ce{C4H4NH})$    & (6e,5o) & (6e,5o)     \\
pyridine $(\ce{C5H5N})$    & (6e,6o) & (8e,7o)  \\
pyrimidine $(\ce{C4H4N2})$ & (6e,6o) & (10e,8o) \\
\hline\hline
\end{tabular}
\vspace{3mm}
\caption{Number of active-space electrons (e) and molecular orbitals (o) for the four species considered in this work.}
\label{tab:active_spaces}
\end{table}

Classical electronic structure simulations for furan, pyrrole, pyridine, and pyrimidine were carried out with an aug-cc-pVDZ basis set at MP2/aug-cc-pVTZ optimized geometries using the PySCF \cite{sun2018pyscf,sun2020recent} and Q-Chem packages~\cite{shao2015advances}.

An active space for each of the molecules was constructed to include $\pi\rightarrow\pi^*$ and $n \rightarrow \pi^*$ excitations, the latter arising from excitation of an electron in a lone pair on the hetero-atom into the $\pi$ virtual space.
The orbitals are illustrated in Figure \ref{fig:active_spaces}, and the sizes of the corresponding active spaces are given in Table \ref{tab:active_spaces}. 
Additional information on the active-space molecular orbitals is given in the Appendix. 

For each of the molecules, the corresponding active space was used for subsequent quantum computing and classical calculations.

\subsubsection{Quantum calculations}

Quantum computing active-space ground-state calculations were carried out using the EF computational package from Ref.~\citenum{entanglement-forging}, interfaced with IBM's quantum computing library \library{Qiskit} \cite{aleksandrowicz2019qiskit}, following which excited-state (QSE) simulations were performed with an in-house extension of the EF package, as documented in Ref.~\citenum{motta2023quantum}. 

EF+QSE was simulated in three ways: (i) using 
\library{Qiskit}'s \library{statevector} library, which models quantum operations using matrix-vector multiplications, and 
assumes an ideal (i.e. decoherence-free) quantum computer and infinitely large statistical samples (i.e. in the absence of shot noise).
(ii) using \library{Qiskit}'s \library{QASM} library, which models quantum operations using matrix-vector multiplications but includes the simulation of quantum measurements by sampling the probability distributions for a finite number of times (or ``shot'', \textit{i.e.,} a single outcome of a single quantum measurement) denoted ``QASM, ideal''. 
(iii) using the \library{Runtime} library of \library{Qiskit}  to interface the code with quantum devices.

\begin{figure*}[t!]
\centering
\includegraphics[width=\textwidth]{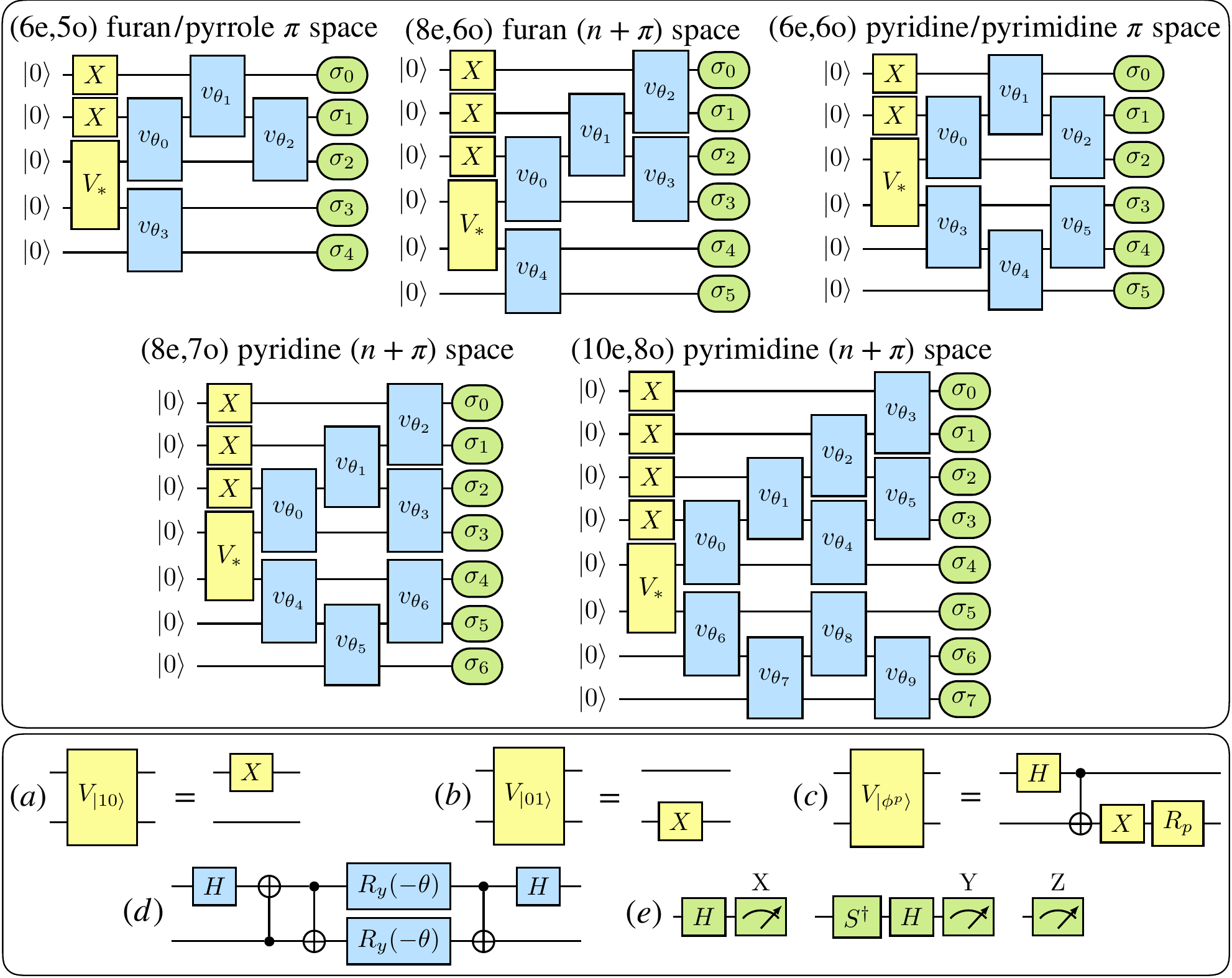}
\caption{Top: quantum circuits simulated in the present work.
Bottom: two-qubit unitaries $V_*$ (yellow) used to bring the initial state $|00 \rangle$ to a state of the form $|10 \rangle$ (a), $|01 \rangle$ (b), and $| \phi^p \rangle$, where the integer $p = 0,1,2,3$ corresponds to the single-qubit gate $R_0 = I$,
$R_1 = ZS$, $R_2 = Z$, $R_3 = S$ (c); compilation of the two-qubit gate $v_{\theta}$ called ``hop-gate'' (blue) into single-qubit and CNOT gates (d); quantum circuits to measure (green) the single-qubit Pauli operators X,Y,Z (e). Additional information can be found in Refs.~\citenum{motta2021emerging} and \citenum{motta2023quantum}.}
\label{fig:ef_circuits}
\end{figure*}

For option (iii), jobs consisting of 150 or 300 circuits and $100,000$ shots for each circuit were submitted on quantum hardware (see below for a description of the circuits). Twirled readout error mitigation \cite{nation2021scalable} was used to mitigate errors arising from readout, dynamical decoupling \cite{viola1998dynamical,kofman2001universal,biercuk2009optimized,rost2020simulation,niu2022effects,niu2022analyzing,ezzell2022dynamical} was used to mitigate errors arising from quantum gates, and post-selection \cite{huggins2021efficient,cohn2021quantum} was used to enforce conservation of electron number.

The hardware simulations were performed on the superconducting quantum devices \device{hanoi} (furan $\pi$-space, $n+\pi$-space, and pyrrole $\pi$-space), \device{geneva} (pyridine $\pi$-space pyrimidine $\pi$-space), \device{cairo} (pyridine $n+\pi$-space) and \device{sherbrooke} (pyrimidine-$n+\pi$ space). 

\begin{table}[t!]
\centering
\begin{tabular}{cccccc}
\hline\hline
system & qubits & var. param. & gates & depth \\
\hline
(6e,5o)  & 5 &  6 & (32,13) & 26 \\
(8e,6o)  & 6 &  7 & (39,16) & 26 \\
(6e,6o)  & 6 &  8 & (42,19) & 26 \\
(8e,7o)  & 7 &  9 & (49,22) & 26 \\
(10e,8o) & 8 & 11 & (64,31) & 32 \\
\hline\hline
\end{tabular}
\vspace{3mm}
\caption{Number of qubits, variational parameters (one for every hop-gate and two corresponding to the Schmidt coefficients), single- and two-qubit gates (4,4,2 single-qubit and 1,3,0 two-qubit gates for each $V$, hop-gate and measurement, respectively) and depth of the circuits shown in Figure \ref{fig:ef_circuits}.}
\label{tab:ef_circuits}
\end{table}

\subsection{Quantum circuits}

The quantum circuits simulated in this study are shown in Figure \ref{fig:ef_circuits}.
In these circuits, qubits are initialized in the state $|0 \rangle$ (corresponding
to an unoccupied orbital).
The yellow blocks marked $X$ denote single-qubit Pauli $X$ gates, which are used to bring
a single qubit in the state $|1 \rangle$ (corresponding to an occupied orbital).

The yellow blocks marked $V$ represent two-qubit unitaries \cite{motta2023quantum}
used to bring the initial state $|00 \rangle$ to a state of the form $|10 \rangle$, $|01 \rangle$, or $| \phi^p_{kl} \rangle$ (see (a), (b), (c) in the bottom panel of the figure). For example, in the case of a (6e,6o) active space, application of the $X$ and $V$ gates prepares qubits in either the state $|1 1 1 0 0 0 \rangle$, or 
$|1 1 0 1 0 0 0 \rangle$, 
or a superposition of such states.
We choose these two computational basis states to highlight entanglement across a pair of active-space orbitals, one occupied and one unoccupied.
The blue blocks, $v_{\theta_i}$, each denote a two-qubit unitary called a hop-gate.
Hop-gates are number-conserving functionally complete 2-qubit gates \cite{eddins2022doubling,motta2023quantum} used to correlate electrons.
Note that, while products of hop-gates have been used as an ansatz for EF simulations in the past,
other choices are also possible, 
such as qubit hardware-efficient circuits \cite{kandala2017hardware,d2022accuracy}
or qubit coupled-cluster \cite{ryabinkin2018qubit}. 
Finally, the green blocks, marked $\sigma_k$, denote the measurement output of a Pauli operator.
Details about the construction of the $V$, $v_{\theta}$, and $\sigma_k$ operations is in panel (d), and more information can be found in Refs.~\citenum{eddins2022doubling,motta2023quantum,motta2021emerging}.
The computational cost of the quantum circuits shown in Figure \ref{fig:ef_circuits} is given
in Table \ref{tab:ef_circuits}. 

\subsubsection{Classical vertical excitation energy calculations}

For purposes of direct comparison between the quantum computing (EF+QSE) and classical electronic results, 
classical calculations were carried out considering all excitations within the active space defined above using the complete active space configuration interaction (CASCI) method. 

Additional classical calculations were performed to determine excitation energies and oscillator strengths using methods that include the effect of dynamic electron correlation through single and/or double excitations from the occupied orbitals into all the unoccupied orbitals (keeping the 1s orbitals frozen). For brevity, we will call these calculations ``full MO space''. Here, we considered configuration interaction singles (CIS) and equation-of-motion coupled-cluster singles and doubles (EOM-CCSD).

\section{Results}

\subsection{Active-space simulations on classical and quantum hardware}

\begin{figure*}[t!]
\centering
\includegraphics[width=\textwidth]{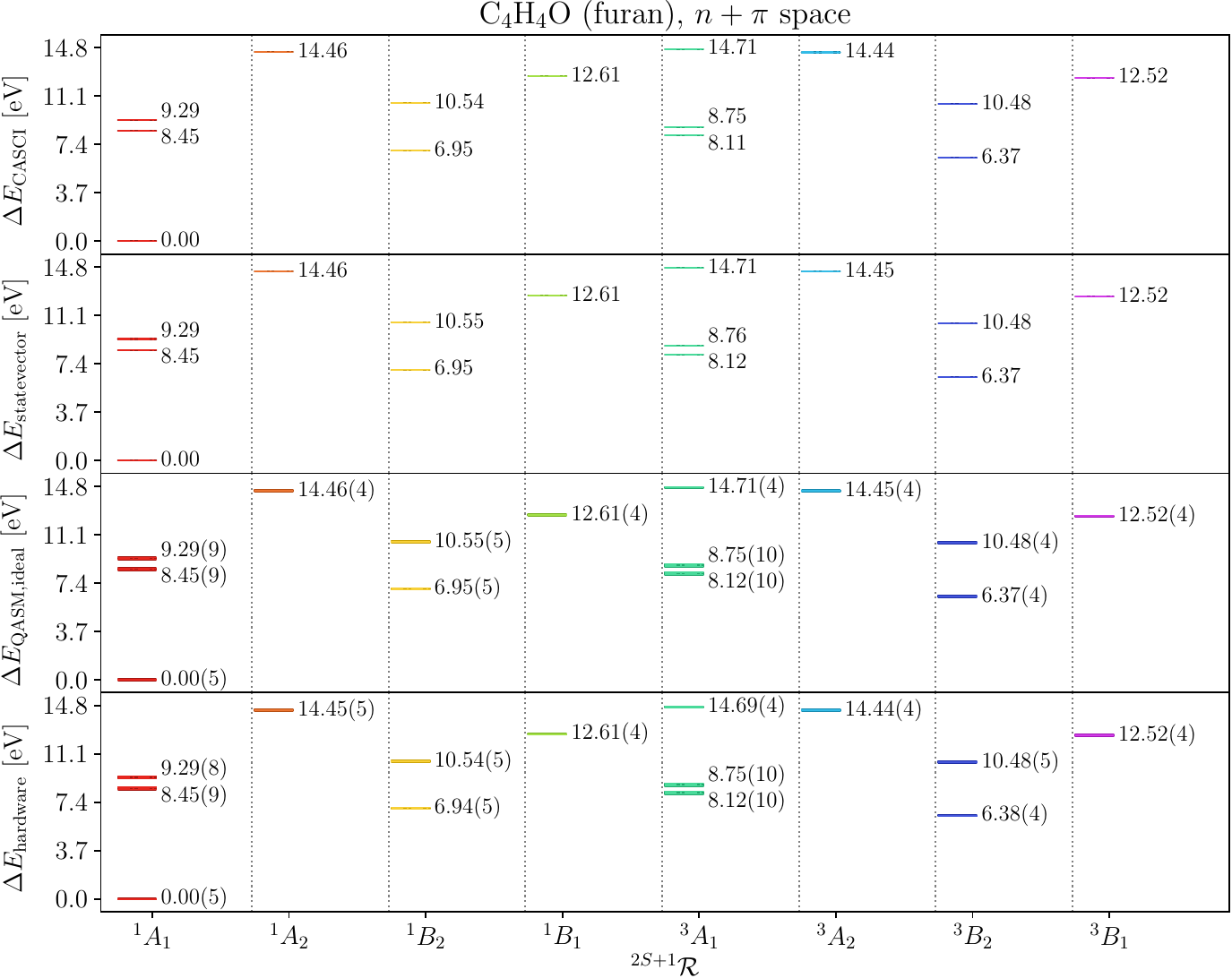}
\caption{Excited-state energies of furan $(\ce{C4H4O})$ in an active space spanned by $\pi$ and one lone-pair, using CASCI and EF+QSE simulated with \library{statevector}, QASM, and hardware (top to bottom).
Energies are measured in eV, and labeled by total spin and irrep of the molecular point group symmetry.}
\label{fig:furan_hardware_act}
\end{figure*}

\begin{figure}[h!]
\centering
\includegraphics[width=0.5\textwidth]{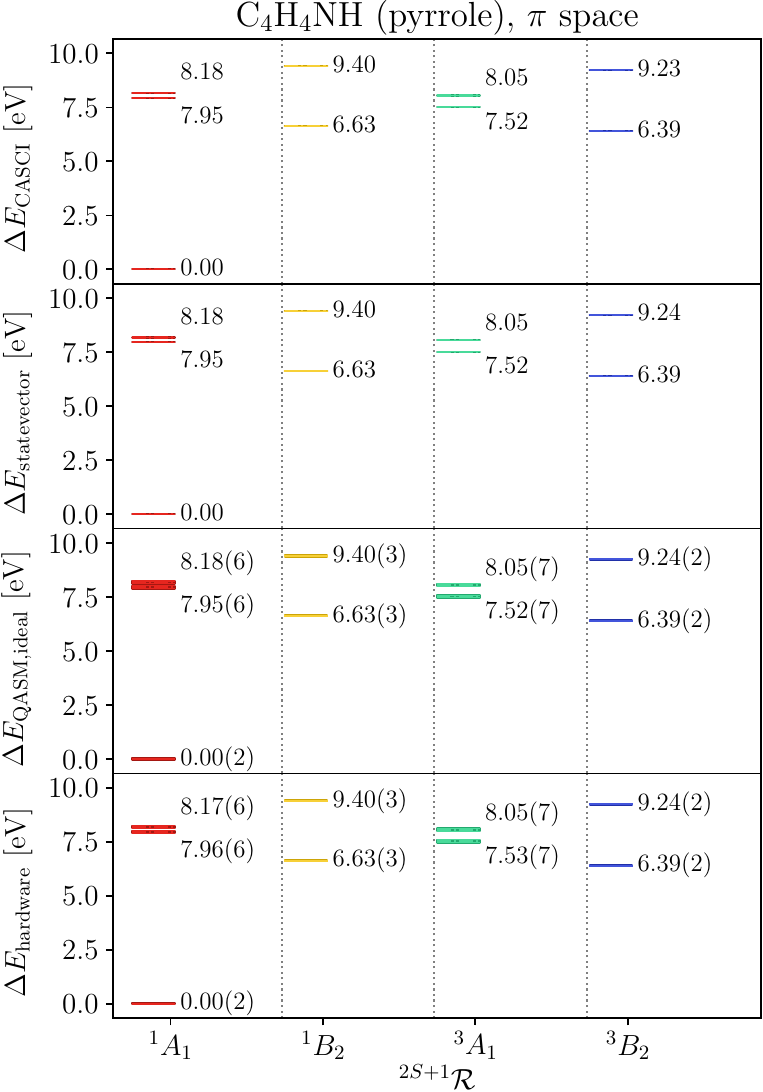}
\caption{Excited-state energies of pyrrole $(\ce{C4H4NH})$ in an active space spanned by $\pi$ orbitals, using CASCI and EF+QSE simulated with \library{statevector}, QASM, and hardware (top to bottom).
Energies are measured in eV, and labeled by total spin and irrep of the molecular point group symmetry.}
\label{fig:pyrrole_hardware_pi}
\end{figure}

\begin{figure*}[t!]
\centering
\includegraphics[width=\textwidth]{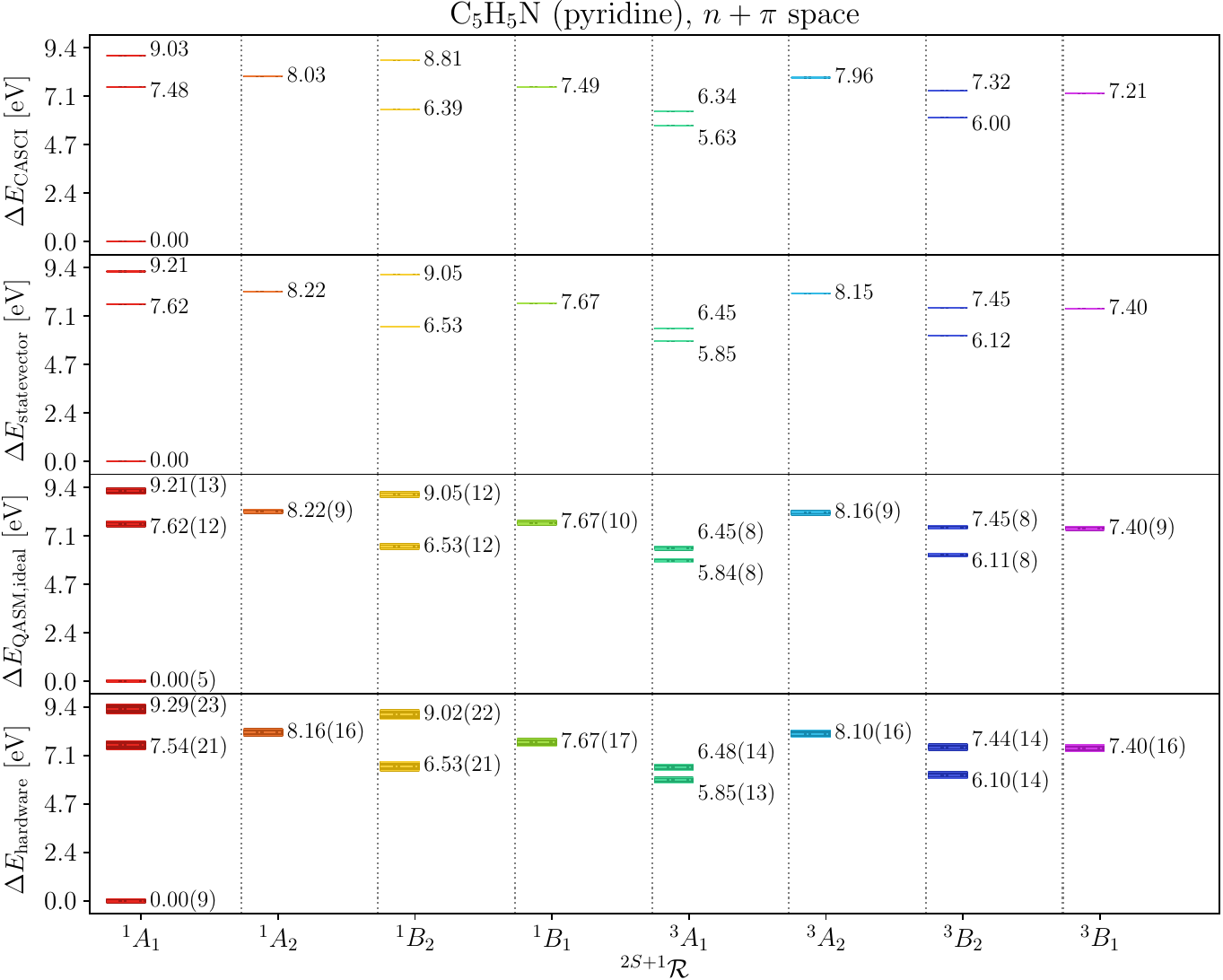}
\caption{Excited-state energies of pyridine $(\ce{C5H5N})$ in an active space spanned by $\pi$ orbitals and a lone-pair, using CASCI and EF+QSE simulated with \library{statevector}, QASM, and hardware (top to bottom).
Energies are measured in eV, and labeled by irrep of the molecular point group symmetry and total spin.}
\label{fig:pyridine_hardware_act}
\end{figure*}

\begin{figure*}[t!]
\centering
\includegraphics[width=\textwidth]{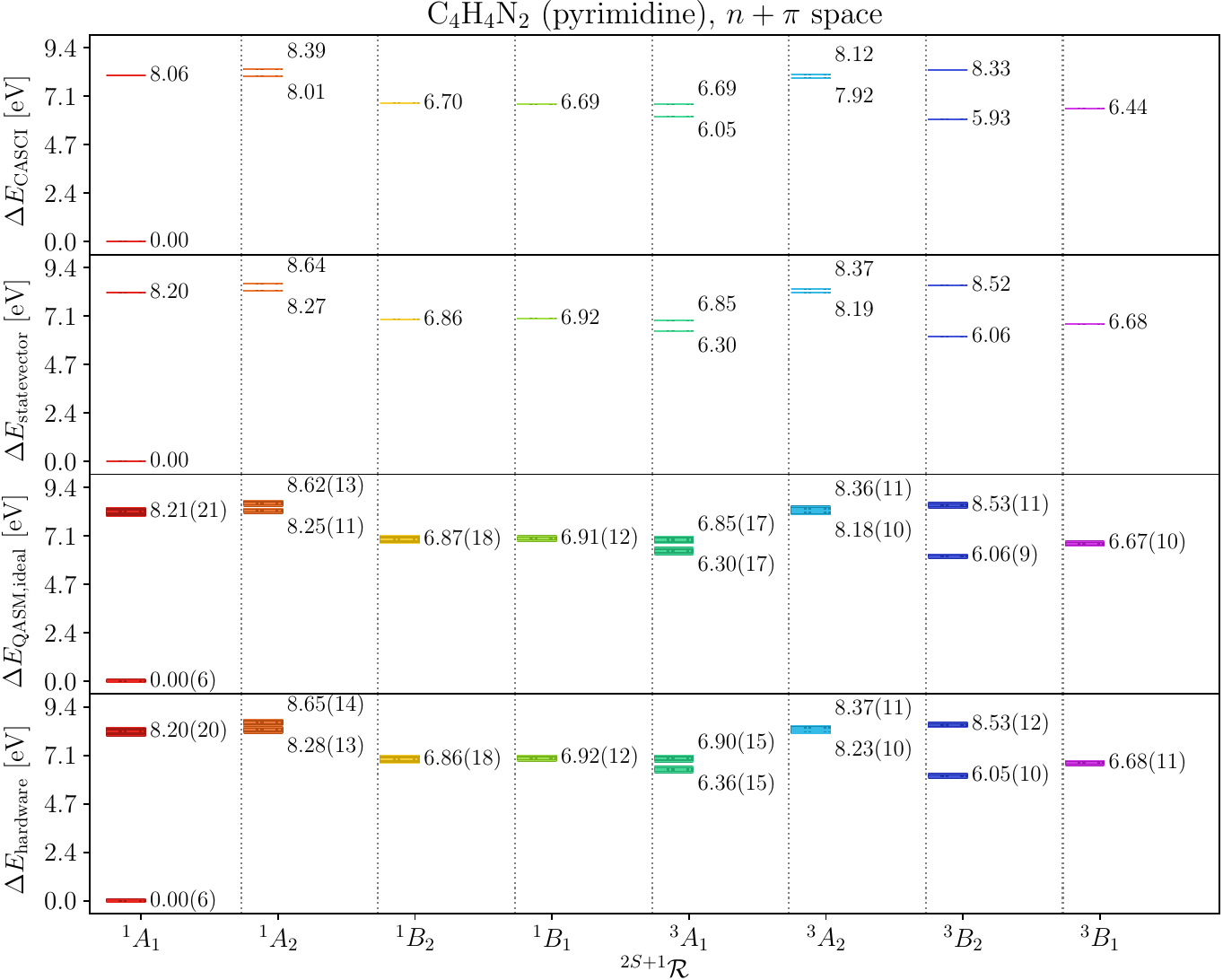}
\caption{Excited-state energies of pyrimidine $(\ce{C4H4N2})$ in an active space spanned by $\pi$ and one lone-pair orbital, using CASCI and EF+QSE simulated with \library{statevector}, QASM, and hardware (top to bottom).
Energies are measured in eV, and labeled by total spin and irrep of the molecular point group symmetry.}
\label{fig:pyrimidine_hardware_act}
\end{figure*}

In Figures \ref{fig:furan_hardware_act}-\ref{fig:pyrimidine_hardware_act} we show the energies for the excited electronic states of furan, pyrrole, pyridine and pyrimidine, respectively, using the active spaces in Fig.~\ref{fig:active_spaces}.
For furan, pyridine and pyrimidine, these active spaces include $\pi$ orbitals and lone-pair MO(s). For completeness, in the Appendix we show the corresponding figures for the $\pi$-space simulations for furan, pyridine and pyrimidine.
For each group of electronic states, only a subset of excited-state energies are shown in order to focus on physically-relevant low-lying excited states. 

\subsubsection{Accuracy and precision}

The figure for each molecule has a top panel, marked as ``CASCI'', 
which shows the excited-state energies from a CASCI calculation.
The second panel from the top, marked as ``statevector'', shows excited-state energies from EF+QSE using an exact noiseless quantum circuit simulator. The comparison between the first and the second panel  highlights the  approximations of the EF+QSE algorithm. 
The third panel, marked ``QASM, ideal''
uses the \library{QASM} library of
\library{Qiskit} to perform noiseless simulations of quantum circuits with a finite number of shots. The difference between the second and the third panel illustrates the impact of statistical uncertainties (which are absent in the first and second panels) in the determination of excited-state energies.
The fourth panel, marked ``hardware'', shows excited-state energies computed from specific quantum hardware (see Methods section) after error mitigation and post-selection techniques. 
The difference between the third and fourth panels shows the impact of the remaining decoherence on the computed energies. The differences between the panels are quantified in Table~\ref{table:sv_versus_casci}. 

In Table~\ref{table:sv_versus_casci} (left) we show the minimum, maximum, and average deviations between CASCI and \library{statevector} excitation energies for the systems studied in this work. We can see that the \library{statevector} overestimates excitation energies by up to 0.3 eV. 

Excited-state energies from \library{statevector} and \library{QASM} are statistically compatible with each other, as we document in Table \ref{table:sv_versus_casci} (middle). Average deviations between
excited-state energies from \library{statevector} and \library{QASM} are of the order of meV, and the chi-squared functional 
\begin{equation}
\chi^2 = \sum_{i=1}^n \frac{ \left( \Delta\Delta E_i \right)^2 }{ n \sigma_i^2 }
\end{equation}
of these energy deviations is below 0.05. We remark, however, that the average statistical uncertainties on \library{QASM} excitation energies are between 0.04 and 0.17 eV. 

Finally, in the right-side of Table \ref{table:sv_versus_casci} we compare \library{QASM} and hardware results. Although excited-state energies from quantum hardware are slightly noisier than their \library{QASM} counterparts, results from quantum hardware are statistically compatible with results from \library{statevector} simulations. Statistical uncertainties are smaller for the 5-membered rings, pyrrole and furan, which is consistent with the fact that in general the average statistical uncertainty increases when the active space is expanded in size. 

\begin{table}[h!]
\setlength{\tabcolsep}{2.5pt}
\centering
\begin{tabular}{l|ccc|ccc|ccc}
\hline\hline
system & \multicolumn{3}{c|}{CASCI vs \library{statevector}} & \multicolumn{3}{c|}{\library{statevector} vs \library{QASM}} & \multicolumn{3}{c}{\library{QASM} vs hardware}\\
 & $\Delta\Delta E_{min}$ & $\Delta\Delta E_{max}$ & $\overline{\Delta\Delta E}$ & $\overline{\Delta\Delta E}$ & $\sigma$ & $\chi^2$ & $\overline{\Delta\Delta E}$ & $\sigma$ & $\chi^2$\\
\hline
\small{furan}      & 0.0003 & 0.0063 & 0.0028 &  0.0015 & 0.0584 & 0.0031 &  0.0056 & 0.0590 & 0.0408 \\
\small{pyrrole}    & 0.0007 & 0.0020 & 0.0014 & -0.0013 & 0.0441 & 0.0010 & -0.0003 & 0.0439 & 0.0047 \\
\small{pyridine}   & 0.1123 & 0.2440 & 0.1693 & -0.0020 & 0.0968 & 0.0022 &  0.0110 & 0.1715 & 0.0495 \\
\small{pyrimidine} & 0.1305 & 0.2707 & 0.2126 &  0.0055 & 0.1340 & 0.0083 & -0.0141 & 0.1347 & 0.0369 \\
\hline\hline
\end{tabular}
\vspace{3mm}
\caption{
Left: Minimum, maximum, and average deviations $\Delta \Delta E$ between CASCI and EF+QSE (\library{statevector}) excitation energies $\Delta E$. Quantities are measured in eV and computed over all spin polarizations, irreps, and excited states in Figures~\ref{fig:furan_hardware_act}-\ref{fig:pyrimidine_hardware_act}.
Middle: average deviation $\overline{\Delta\Delta E}$ between EF+QSE excited-state energies computed with \library{statevector} and \library{QASM} (left), average uncertainty $\sigma$ on EF+QSE excited-state energies computed with \library{QASM} (middle), and $\chi^2$ for deviations between EF+QSE excited-state energies computed with \library{statevector} and \library{QASM}. Quantities are measured in eV (except $\chi^2$, which is dimensionless).
Right: same as the middle, but comparing \library{QASM} with quantum hardware.
}
\label{table:sv_versus_casci}
\end{table}

\subsection{Comparison with classical calculations and experimental data}

\begin{table*}[!htbp]
\setlength{\tabcolsep}{1.5pt}
\centering
\begin{tabular}{*9c}
\hline\hline
State &	CASCI	& \library{Sv} &	hardware 	&	CIS	&	EOM &	Best Est. & Expt.	& $f$ \\
\hline
1$^3B_2$ \small{($\pi\rightarrow\pi^*$)}	&	6.37	&	6.37	&	6.38(4)	&	3.31	&	4.11	& 4.17 &	4.02$^a$; 3.99$^b$	&	0.0000\\
1$^3A_1$ \small{($\pi\rightarrow\pi^*$)}	& 8.11	&	8.11	&	8.12(10)	&	4.90	&	5.48	& 5.48 &	5.22$^a$	&	0.0000 \\
2$^3A_1$ \small{($\pi\rightarrow\pi^*$)}	&	8.75	&	8.76	&	8.75(10)	&	7.05	&	6.83	&   &	$\sim$6.5$^b$ &	0.0000\\
2$^3B_2$ \small{($\pi\rightarrow\pi^*$)} 	&	10.48	&	10.48	&	10.48(5) &	6.97	&	7.17	&   &	$\sim$6.5$^b$	&	0.0000\\
[2mm]
1$^1B_2$ \small{($\pi\rightarrow\pi^*$)} &	6.95	&  6.95	&	6.94(5)	&	6.30	&	6.46	&	6.32 & 
6.04$^b$	&	0.1794\\
2$^1A_1$ \small{($\pi\rightarrow\pi^*$)}	&	8.45 &	8.45	&	8.45(9)	&	7.38	&	6.79	& 6.57 &	5.8$^b$	&	0.0000\\
3$^1A_1$ \small{($\pi\rightarrow\pi^*$)}	&	9.29 &	9.29	&	9.29(8)	&	8.68	&	8.32	&	8.13 & 
7.8$^b$	&	0.4010\\
2$^1B_2$  \small{($\pi\rightarrow\pi^*$)} &	10.54 &	10.54	&	10.54(5)	&	8.82	&	9.00	&   &	8.7$^b$	&	0.1761 \\
\hline\hline
\end{tabular}
\vspace{3mm}
\caption{Excitation energies for the triplet and singlet low-lying states of furan and corresponding oscillator strengths, $f$. Results from the present work are compared with CIS and EOM-CCSD at aug-cc-pVDZ level of theory, ``best estimates'' from Ref.~\citenum{schreiber2008benchmarks}, and with experiments ($a$, $b$ for Refs.~\citenum{flicker1976electron} and \citenum{Palmer1995}). CASCI corresponds to the results with an active space of $\pi$ and one lone-pair orbital. $\mathsf{Sv}$ abbreviates $\mathsf{statevector}$.}
\label{table:experimentFuran}
\end{table*}

\begin{table*}[!htbp]
\setlength{\tabcolsep}{2.0pt}
\centering
\begin{tabular}{*9c}
\hline\hline
State &	CASCI	& \library{Sv} &	hardware 	&	CIS	&	EOM &	Best Est. & Expt.	& $f$ \\
\hline
1$^3B_2$ \small{($\pi\rightarrow\pi^*$)}	&	6.39	&	6.39	&	6.39(2)	&	3.73	&	4.42	&	4.48 & 4.21$^{a,b}$	&	0.0000\\
1$^3A_1$ \small{($\pi\rightarrow\pi^*$)}	&	7.52	&	7.52	&	7.53(7)	&	5.16	&	5.50	& 5.51 &	5.10$^{b}$	&	0.0000\\
2$^3A_1$ \small{($\pi\rightarrow\pi^*$)}	&	8.05	&	8.05	&	8.05(7)	&	6.18	&	6.29	&  &		&	0.0000\\
2$^3B_2$ \small{($\pi\rightarrow\pi^*$)} 	&	9.23	&	9.24	&	9.24(2)	&	6.49	&	6.58	&  &	{}	&	0.0000\\
[2mm]
1$^1B_2$ \small{($\pi\rightarrow\pi^*$)} 	&	6.63	&	6.63	&	6.63(3)	&	6.32	&	6.32	& 6.57 &	5.98$^{a}$	&	0.1831\\
2$^1A_1$ \small{($\pi\rightarrow\pi^*$)}	&	7.95	&	7.95	&	7.96(6)	&	7.34	&	6.48	& 6.37 & 	&	0.0004\\
2$^1B_2$  \small{($\pi\rightarrow\pi^*$)}	&	9.40	&	9.40	&	9.40(3)	&	7.07	&	7.05	& {} &	{}	&	0.0000\\
3$^1A_1$ \small{($\pi\rightarrow\pi^*$)}	&	8.18	&	8.18	&	8.17(6)	&	7.84	&	7.83	& 7.91 & 7.54$^{a}$	&	0.3981\\
\hline\hline
\end{tabular}
\vspace{3mm}
\caption{Excitation energies for the triplet and singlet low-lying states of pyrrole and corresponding oscillator strengths, $f$. Results from the present work are compared with CIS and EOM-CCSD at aug-cc-pVDZ level of theory, ``best estimates'' from Ref.~\citenum{schreiber2008benchmarks}, and with experiments ($a$, $b$ for Refs.~\citenum{flicker1976electron} and \citenum{van1976triplet}). CASCI corresponds to the results with an active space of $\pi$ and one lone-pair orbital. $\mathsf{Sv}$ abbreviates $\mathsf{statevector}$.}
\label{table:experimentPyrrole}
\end{table*}

\begin{table*}[!htbp]
\setlength{\tabcolsep}{1.2pt}
\centering
\begin{tabular}{*9c}
\hline\hline
State &	CASCI	& \library{Sv} &	hardware 	&	CIS	&	EOM &	Best Est. & Expt.	& $f$ \\
\hline
1$^3A_1$ \small{($\pi\rightarrow\pi^*$)}	&	5.63	&	5.85	&	5.85(13)	&	3.54	&	4.15	& 4.06 &	$\sim$4.1$^{a,b}$; 3.86$^c$	&	0.0000\\
1$^3B_2$	\small{($\pi\rightarrow\pi^*$)} &	6.00	&	6.12	&	6.10(14)	&	4.65	&	4.57 & 4.64	&	$\sim$4.84$^{a,b}$; 4.47$^c$	&	0.0000\\
2$^3A_1$ \small{($\pi\rightarrow\pi^*$)}	&	6.34	&	6.45	&	6.48(14)	&	5.02	&	5.11	& 4.91 &	$\sim$4.84$^{a,b}$	&	0.0000\\
2$^3B_2$ \small{($\pi\rightarrow\pi^*$)} 	&	7.32	&	7.45	&	7.44(14)	&	6.16	&	6.28	& 6.08 &	6.09$^c$	&	0.0000\\
\hline
1$^3B_1$	\small{($n\rightarrow\pi^*$)} &	7.21	&	7.40	&	7.40(16)	&	5.10	&	4.85	& 4.25 &	$\sim$4.1$^a$; 4.12$^c$	&	0.0000\\
1$^3A_2$	\small{($n\rightarrow\pi^*$)} &	7.96	&	8.15	&	8.10(16)	&	7.13	&	5.55	& 5.28 &	$\sim$5.43$^{a,b}$; 5.40$^c$ &	0.0000\\
[2mm]
1$^1B_2$ \small{($\pi\rightarrow\pi^*$)} 	&	6.39	& 	6.53	&	6.53(21)	&	6.10	&	5.29 & 	4.85 &	4.99$^{a,c}$	&	0.0301\\
2$^1A_1$ \small{($\pi\rightarrow\pi^*$)}	&	7.48	&	7.62	&	7.54(21)	&	6.45	&	6.73	& 6.26 &	6.38$^b$	&	0.0021\\
2$^1B_2$  \small{($\pi\rightarrow\pi^*$)}	&	8.81	&	9.05	&	9.02(22)	&	7.96	&	7.55	& 7.27 &	7.22$^b$; 7.20$^c$	&	0.5377\\
3$^1A_1$ \small{($\pi\rightarrow\pi^*$)}	&	9.03	&	9.21	&	9.29(23)	&	8.22	&	7.76	& 7.18 &	7.22$^b$; 6.39$^c$	&	0.4844\\
\hline
1$^1B_1$ \small{($n\rightarrow\pi^*$)}	&	7.49	&	7.67	&	7.67(17)	&	6.14	&	5.18	& 4.59	& 4.44$^b$; 4.78$^c$	&	0.0051\\
2$^1A_2$ \small{($n\rightarrow\pi^*$)}	&	8.03	&	8.22	&	8.16(16)	&	7.38	&	5.61	& 5.11 &	5.43$^b$; 5.40$^c$	&	0.0000\\
\hline\hline
\end{tabular}
\vspace{3mm}
\caption{Excitation energies for the triplet and singlet low-lying states of pyridine and corresponding oscillator strengths, $f$. Results from the present work are compared with CIS and EOM-CCSD at aug-cc-pVDZ level of theory, ``best estimates'' from Ref.~\citenum{schreiber2008benchmarks}, and with experiments ($a$, $b$, $c$ for Refs.~\citenum{Cai2000},~\citenum{Walker1990}, and \citenum{linert2016electron}). CASCI corresponds to the results with an active space of $\pi$ and one lone-pair orbital. $\mathsf{Sv}$ abbreviates $\mathsf{statevector}$.}
\label{table:experimentPyridine}
\end{table*}

\begin{table*}[!htbp]
\setlength{\tabcolsep}{1.5pt}
\centering
\begin{tabular}{*9c}
\hline\hline
State &	CASCI	& \library{Sv} &	hardware 	&	CIS	&	EOM &	Best Est. & Expt.	& $f$ \\
\hline
1$^3A_1$ \small{($\pi\rightarrow\pi^*$)}	& 	6.05	&	6.30	&	6.36(15)	&	3.84	&	4.38 &	{} &	4.42$^a$	&	0.0000\\
1$^3B_2$ \small{($\pi\rightarrow\pi^*$)}	& 	6.44	&	6.68	&	6.68(11)	&	4.77	&	5.02	& {} &	4.93$^a$; 5.05$^b$	&	0.0000\\
2$^3A_1$ \small{($\pi\rightarrow\pi^*$)}	& 	6.69	&	6.85	&	6.90(15)	&	5.53	&	5.44	& {} &	{}	&	0.0000\\
2$^3B_2$ \small{($\pi\rightarrow\pi^*$)} & 		9.44	&	9.71	&	9.75(9)	&	6.52	&	6.60	& {} &	{}	&	0.0000\\
\hline
1$^3B_1$	 \small{($n\rightarrow\pi^*$)} & 5.93	&	6.06	&	6.05(10)	&	5.04	&	4.21	& {} &	3.85$^a$; 3.6$^b$	&	0.0000\\
1$^3A_2$	\small{($n\rightarrow\pi^*$)} & 	7.92	&	8.19	&	8.23(10)	&	5.71	&	4.79	& {} &	4.18$^a$; $\sim$4.5$^b$	&	0.0000\\
[2mm]
1$^1B_2$ \small{($\pi\rightarrow\pi^*$)} & 		6.69	&	6.92	&	6.92(12)	&	6.43	&	5.52	& 5.44 &	5.12$^b$; 5.18$^a$	&	0.0293\\
2$^1A_1$ \small{($\pi\rightarrow\pi^*$)}	& 	8.06	&	8.20	&	8.20(20)	&	6.86	&	7.00	& 6.95 &	6.70$^b$	&	0.0277\\
3$^1A_1$ \small{($\pi\rightarrow\pi^*$)}	& 	9.41	&	9.61	&	9.69(20)	&	8.35	&	7.80	& {} &	$\sim$7.6$^b$	&	0.4289\\
2$^1B_2$ \small{($\pi\rightarrow\pi^*$)}	& 	9.56	&	9.83	&	9.87(9)	&	8.75	&	8.10 &	{} &	$\sim$7.6$^b$	&	0.3346\\
\hline
1$^1B_1$ \small{($n\rightarrow\pi^*$)}	& 	6.70	&	6.86	&	6.86(18)	&	5.87	&	4.63	&	4.55 & 4.18$^a$	&	0.0064\\
1$^1A_2$ \small{($n\rightarrow\pi^*$)} & 		8.01	&	8.27	&	8.28(13)	&	6.54	&	5.04	& 4.91 & $\sim$4.7$^b$; 4.69$^a$	&	0.0000\\
2$^1A_2$  \small{($n\rightarrow\pi^*$)}		 & 8.39	&	8.64	&	8.65(14)	&	7.48	&	6.19	& {} &	$\sim$5.7$^b$; 5.67$^a$	&	0.0000\\
2$^1B_1$ \small{($n\rightarrow\pi^*$)}	& 	9.56	&	9.83	&	9.87(2)	&	7.74	&	6.51 &	{} &	$\sim$6.0$^b$; 6.02$^a$	&	0.0000\\
\hline\hline
\end{tabular}
\vspace{3mm}
\caption{Excitation energies for the triplet and singlet low-lying states of pyrimidine and corresponding oscillator strengths, $f$. Results from the present work are compared with CIS and EOM-CCSD at aug-cc-pVDZ level of theory, ``best estimates'' from Ref.~\citenum{schreiber2008benchmarks}, and with experiments ($a$, $b$ for Refs.~\citenum{linert2015study},~\citenum{palmer1990electronic}). CASCI corresponds to the results with an active space of $\pi$ and one lone-pair orbital. $\mathsf{Sv}$ abbreviates $\mathsf{statevector}$.}
\label{table:experimentPyrimidine}
\end{table*}

In Tables \ref{table:experimentFuran}-\ref{table:experimentPyrimidine} we compare the energies calculated in this study for the CASCI, \library{statevector} and hardware simulations, with energies calculated using electronic structure methods in the full MO space (see Methods), as well as theoretical and experimental reference data.
We note that oscillator strengths for transitions of $A_2$ symmetry are zero because no component of the dipole operator transforms as $A_2$. Transitions of $B_1$ symmetry have zero oscillator strengths unless lone-pair orbitals are included in the active space, as excitations from the $A_1$ and $B_2$ occupied lone-pair(s) are the only ones that can be coupled to $\pi^*$ excited states (through the component of the dipole operator that transforms as $B_1$).

As some transitions do not have available experimental energies,  EOM-CCSD/aug-cc-pVDZ data and, when available, values from  Ref.~\citenum{schreiber2008benchmarks} are presented as a reference for the calculated energies. Ref.~\citenum{schreiber2008benchmarks} presents data from highly-correlated electronic structure methods with large basis sets, including CC3 (iterative coupled cluster singles and doubles with connected triple excitations~\cite{koch1997cc3}) and CASPT2. These methods give results that are close to the available experimental data. 

CIS results are in significantly worse agreement with experiments than EOM-CCSD results, differing from experimental data by approximately 0.5-1 eV (on average) due to neglect of electron correlation effects. In contrast, EOM-CCSD excitation energies are within 0.5 eV of most reported experimental energies. In our active-space calculations, we can see that the excitation energies are overestimated, for all four molecules, compared to experiments. Discrepancies are significant in the triplet and $n\rightarrow\pi^*$ transitions. For example, in the case of pyrrole and furan, the lowest triplet excitation (1$^3B_2$) is overestimated by $\sim 2$ eV.
In addition, it is worth noting that our results generally preserve the ordering of the transition energies when compared to the EOM-CCSD calculations and experimental data, with the only exceptions being the 1$^1B_2$ and 2$^1A_1$ excitations in furan. 

\section{Discussion}

We will now interpret the results in light of the challenges outlined in the Introduction.

(i) impact of approximations: the discrepancies between CASCI and EF+QSE calculations arise from the approximations in EF+QSE.
First, EF is a heuristic method that approximates the ground state of a Hamiltonian as a linear combination of tensor products, as shown in Eq.~\eqref{eq:ef_target}. These are obtained applying a parameterized quantum circuit (ansatz) to a computational basis state. In this work, we employed a product of hop-gates (see ``Calculation Details'' section) as ansatz, and two computational basis states. This is clearly an approximation. Second, QSE makes use of single and double electronic excitations, see Eq.~\eqref{eq:qse_basis}. This approach is akin to multireference CISD, a method known to be neither rigorously size-extensive nor size-intensive.

The discrepancies between active-space calculations (from CASCI to hardware) and experimental data result from the active-space approximation employed in this work, which neglects dynamical electronic correlation. This conclusion finds support in: (a) the good agreement between CASCI and EF+QSE active-space calculations, despite the approximations of EF+QSE, (b) the sizable differences between our active-space calculations and full MO space calculations (EOM-CCSD and ``best estimates''), and (c) the better agreement between full MO space calculations and experimental data. The latter calculations can account for dynamical electron correlation, an important factor in obtaining accurate excited-state energies.

(ii) statistical uncertainties: the excited-state energies presented in this work are accompanied by statistical uncertainties. These uncertainties do not stem from noise affecting quantum hardware (in fact, error bars of similar magnitude are observed in \library{QASM} results) but from the probabilistic nature of quantum measurements. While statistical uncertainties are acceptable for the present study, they increase as the active-space size increases. This may pose a significant limitation in calculations employing larger active spaces.

(iii) quantum hardware: hardware calculations agree satisfactorily with CASCI and \library{statevector} calculations. However, QSE is a measurement-intensive method, as it requires to evaluate the expectation value of order $(N_o N_v N)^4$ fermionic operators, see Eqs.~\eqref{eq:qse_equation} and \eqref{eq:qse_basis}, where $N_o/N_v$ are the active-space occupied/virtual orbitals and $N=N_o+N_v$. The high measurement cost of QSE, combined with the rate at which information can be extracted from a quantum computer (e.g. approximately $1 \mu s$ on superconducting devices), is a limiting factor for larger calculations.

\section{Conclusions}

In this work, we have demonstrated the combination of EF and QSE methods to simulate excited states of molecular systems within an active space. We applied EF+QSE to four organic molecules representative of moieties found in biological systems (furan, pyrrole, pyridine, and pyrimidine). Focusing on active spaces of 5 to 8 $\pi$ and nonbonding orbitals, we simulated excited states on classical computers and IBM quantum hardware, using 5 to 8 qubits and error mitigation techniques.

This work was motivated by challenges in calculating excited-state energies on today's quantum hardware, as discussed in the Introduction and Discussion sections, and indicates directions for future work to address these challenges.

(i) impact of approximations: first, we described and quantified the approximations in EF+QSE. EF can be improved by exploring ansatzes different from the hopgates used here~\cite{ryabinkin2018qubit}. QSE can be improved by modifying the subspace where the Schr\"{o}dinger equation is projected, e.g. through similarity transformations \cite{asthana2023quantum}.

Even more importantly, our simulations are restricted to an active space to demonstrate quantum computing methods for near-term devices. Therefore, they neglect dynamical electronic correlation associated with transitions to/from orbitals outside the active space. This limitation can be mitigated by using quantum computing algorithms as active-space solvers within hybrid quantum-classical computing algorithms. These approaches include orbital-optimized simulations~\cite{tilly2021reduced}, akin to classical complete active-space self-consistent field (CASSCF), and methods that account for dynamical correlation, e.g. CASPT2 \cite{tammaro2022n}.

(ii) statistical uncertainties: quantum hardware results are affected by error bars due to the quantum-mechanical measurement of QSE matrix elements and their propagation to excited-state energies through the solution of the GEEV Eq.~\eqref{eq:qse_equation}. While statistical uncertainties can be reduced by simply increasing the number of shots, it is also important to explore variance-reduction techniques.

(iii) quantum hardware: while hardware results reported here are in good agreement with exact \library{statevector} simulations of EF+QSE, implementing these methods requires a considerable overhead of measurements. Reducing the number of measurements, e.g. by using optimization techniques \cite{choi2023measurement}, is an important effort for achieving more practical simulations.

Finally, this work includes, to the best of our knowledge, the largest reported calculation of electronic excited states on a quantum computer (8 qubits) describing a chemical system with 8 spatial-orbitals / 16 spin-orbitals (c.f. Refs.~\citenum{colless2018computation,gao2021applications,huang2022variational,motta2023quantum} using 2 to 6 qubits).
These calculations are also among the most resource-intensive ground-state electronic structure simulations carried out on quantum hardware to date (e.g. Refs. \citenum{rice2021quantum,eddins2022doubling,kandala2017hardware,o2022purification,google2020hartree,zhao2022orbital,guo2022scalable} report variational ground-state simulations using 4 to 12 qubits and Ref. \citenum{huggins2022unbiasing} up to 16 qubits and a hybrid quantum-classical auxiliary-field quantum Monte Carlo \cite{motta2018ab}). 

Our study provides a contribution towards characterizing the performance and limitations of quantum computing methods, and ultimately towards more accurate and realistic quantum simulations of electronic structure, particularly in the context of organic chemistry and biochemistry.

\section*{Acknowledgements}

We gratefully acknowledge Jake Lishman, Doug McClure, Paul Nation, Pedro Rivero-Ramirez, Matt Riedemann, and Jessie Yu for generous help and guidance in carrying out simulation on quantum hardware.

\section*{Dedication}

This paper is dedicated to the memory of Dr. Timothy J. Lee (1959-2022), a brilliant theoretical chemist whose outstanding research included novel algorithms and combined unique sensitivity for numerical methods with profound theoretical insights into the nature of astrochemical molecules and electronic wavefunctions.

\appendix

\section{Active-space orbitals}

The indices, occupancies, energies, characters, and irreducible representations of the active-space MOs are listed in Table \ref{tab:all_active_space}. The irreducible representations are those of C$_{2v}$ molecular point group symmetry.

\begin{table}[h!]
\centering
\begin{tabular}{*{11}{c}}
\hline\hline
\multicolumn{3}{c}{orbital} & \multicolumn{2}{c}{furan} & \multicolumn{2}{c}{pyrrole} & \multicolumn{2}{c}{pyridine} & \multicolumn{2}{c}{pyrimidine} \\

irrep & occ & type & idx & energy & idx & energy & idx & energy & idx & energy\\
\hline
$\ce{b1}$ & 2 & $\pi$  & 12 & -17.251& 14 &-15.428& 17 &-14.717 & 17 & -15.698\\
$\ce{b2}$ & 2 & $n$    & --- & ---   & --- & ---  & --- & ---   & 20 & -11.352\\
$\ce{a1}$ & 2 & $n$    & 16 & -14.737& --- & ---  & 19 & -11.408& 18 & -12.928\\
$\ce{b1}$ & 2 & $\pi$  & 17 & -10.837& 17 & -9.433& 20 & -10.440& 21 & -10.302\\
$\ce{a2}$ & 2 & $\pi$  & 18 & -8.664 & 18 & -8.056& 21 & -9.466 & 19 & -11.503\\

$\ce{b1}$ & 0 & $\pi^*$& 23 &  2.609 & 24 & 2.731 & 27 & 2.361  & 28 & 2.717\\
$\ce{a2}$ & 0 & $\pi^*$& 29 &  3.851 & 29 & 3.929 & 28 & 2.845  & 26 & 2.135\\
$\ce{b1}$ & 0 & $\pi^*$& --- &   --- & --- & ---  & 51 & 8.873  & 48 & 8.441\\
\hline\hline
\end{tabular}
\vspace{3mm}
\caption{Irreducible representations, energies (eV), occupancies and classification of the Hartree-Fock MOs included in the active space of furan $(\ce{C4H4O})$, Pyrrole $(\ce{C4H4NH})$, Pyridine $(\ce{C5H5N})$ and Pyrimidine $(\ce{C4H4N2})$.}
\label{tab:all_active_space}
\end{table}

\section{$\pi$-space simulations}

\begin{figure}[h!]
\centering
\includegraphics[width=0.5\textwidth]{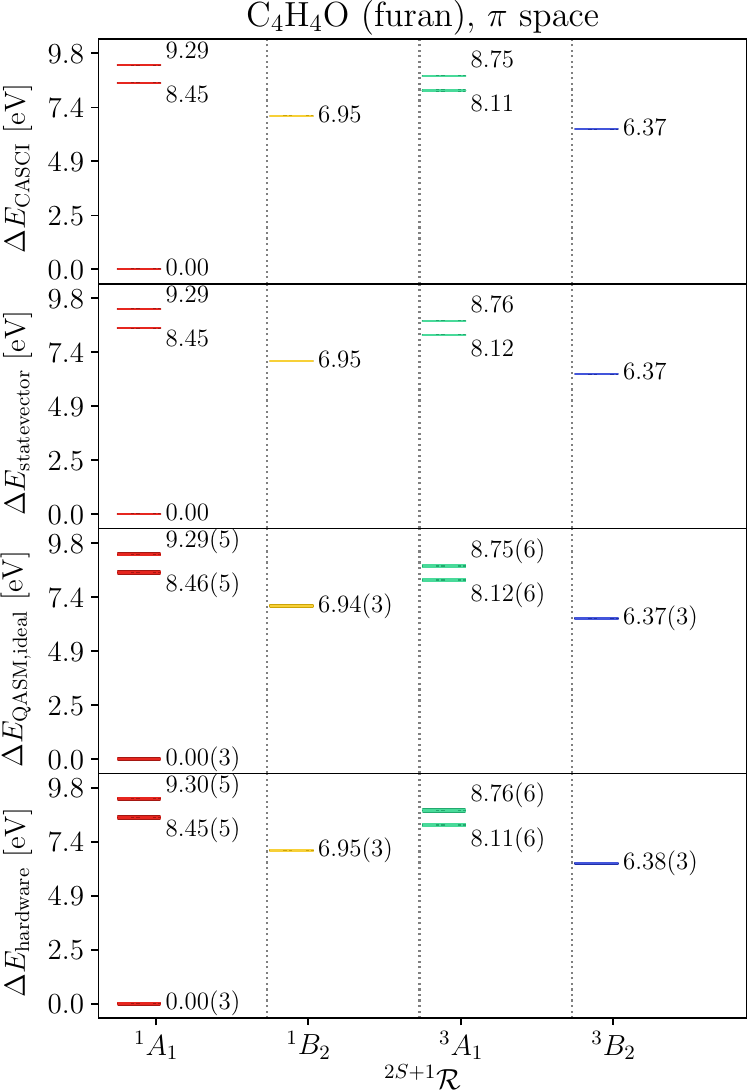}
\caption{Excited-state energies of furan $(\ce{C4H4O})$ in an active space spanned by $\pi$ orbitals, using CASCI and EF+QSE simulated with \library{statevector}, QASM, and hardware (top to bottom).
Energies are measured in eV, and labeled by total spin and irrep of the molecular point group symmetry.}
\label{fig:furan_hardware_pi}
\end{figure}

\begin{figure}[h!]
\centering
\includegraphics[width=0.5\textwidth]{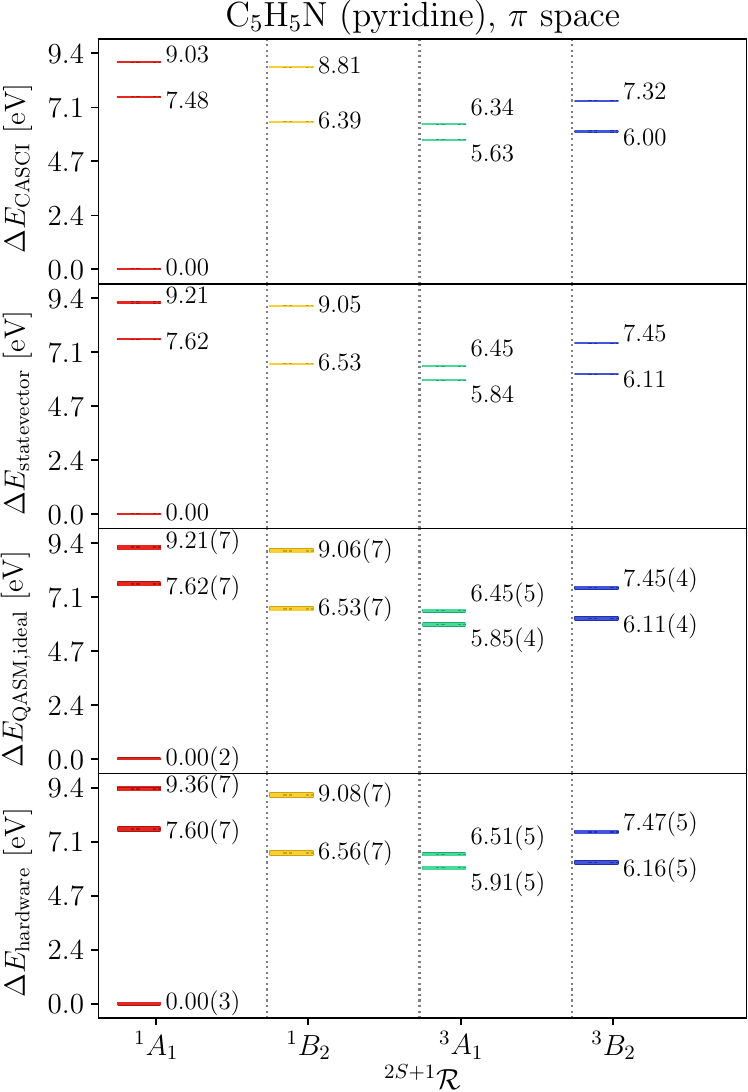}
\caption{Excited-state energies of pyridine $(\ce{C5H5N})$ in an active space spanned by $\pi$ orbitals, using CASCI and EF+QSE simulated with \library{statevector}, QASM, and hardware (top to bottom).
Energies are measured in eV, and labeled by total spin and irrep of the molecular point group symmetry.}
\label{fig:pyridine_hardware_pi}
\end{figure}

\begin{figure}[h!]
\centering
\includegraphics[width=0.5\textwidth]{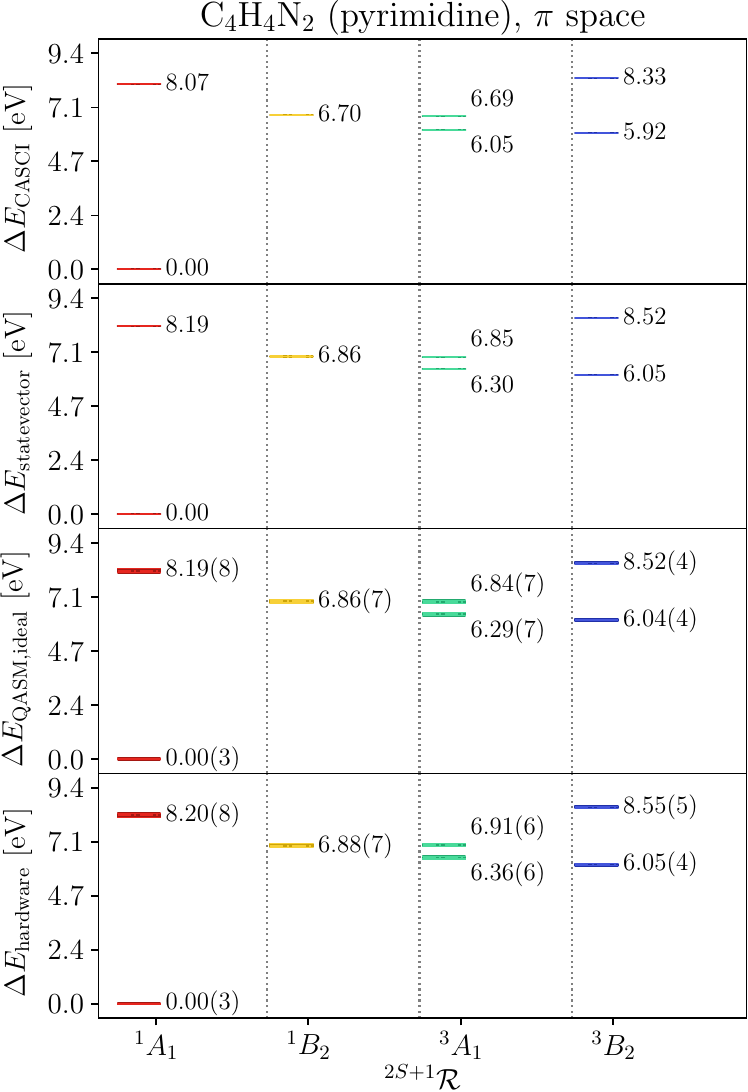}
\caption{Excited-state energies of pyrimidine $(\ce{C4H4N2})$ in an active space spanned by $\pi$ orbitals, using CASCI and EF+QSE simulated with \library{statevector}, QASM, and hardware (top to bottom).
Energies are measured in eV, and labeled by total spin and irrep of the molecular point group symmetry.}
\label{fig:pyrimidine_hardware_pi}
\end{figure}

In Figures~\ref{fig:furan_hardware_pi}, \ref{fig:pyridine_hardware_pi}, and \ref{fig:pyrimidine_hardware_pi}, we show $\pi$-space simulations for furan, pyridine, and pyrimidine respectively. We also report, in Table~\ref{table:sv_versus_casci_appendix}, (i) the minimum, maximum, and average deviations between CASCI and \library{statevector} excitation energies,
(ii) the average deviations between \library{statevector} and \library{QASM} excitation energies, and (iii) the average deviations between \library{QASM} and hardware excitation energies (similar to Table~\ref{table:sv_versus_casci} of the main text).

\begin{table}[h!]
\setlength{\tabcolsep}{2.5pt}
\centering
\begin{tabular}{l|ccc|ccc|ccc}
\hline\hline
system & \multicolumn{3}{c|}{CASCI vs \library{statevector}} & \multicolumn{3}{c|}{\library{statevector} vs \library{QASM}} & \multicolumn{3}{c}{\library{QASM} vs hardware}\\
 & $\Delta\Delta E_{min}$ & $\Delta\Delta E_{max}$ & $\overline{\Delta\Delta E}$ & $\overline{\Delta\Delta E}$ & $\sigma$ & $\chi^2$ & $\overline{\Delta\Delta E}$ & $\sigma$ & $\chi^2$\\
\hline
\small{furan}        & 0.0004 & 0.0054 & 0.0027 &  0.0014 & 0.0456 & 0.0100 & -0.0004 & 0.0465 & 0.0187 \\
\small{pyridine}     & 0.1114 & 0.2420 & 0.1589 & -0.0016 & 0.0568 & 0.0045 & -0.0475 & 0.0603 & 1.1368 \\
\small{pyrimidine}   & 0.1246 & 0.2533 & 0.1688 &  0.0041 & 0.0609 & 0.0080 & -0.0318 & 0.0604 & 0.4244 \\
\hline\hline
\end{tabular}
\vspace{3mm}
\caption{
Left: Minimum, maximum, and average deviations $\Delta \Delta E$ between CASCI and EF+QSE (\library{statevector}) excitation energies $\Delta E$. Quantities are measured in eV and computed over all spin polarizations, irreps, and excited states in Figures~\ref{fig:furan_hardware_pi}-\ref{fig:pyrimidine_hardware_pi}.
Middle: average deviation $\overline{\Delta\Delta E}$ between EF+QSE excited-state energies computed with \library{statevector} and \library{QASM} (left), average uncertainty $\sigma$ on EF+QSE excited-state energies computed with \library{QASM} (middle), and chi-squared for deviations between EF+QSE excited-state energies computed with \library{statevector} and \library{QASM}. Quantities are measured in eV.
Right: same as the middle table, with quantum hardware compared with \library{QASM}.
}
\label{table:sv_versus_casci_appendix}
\end{table}

\pagebreak
\newpage

$ $

\pagebreak
\newpage

$ $

\pagebreak
\newpage


\begin{thebibliography}{71}
\providecommand{\url}[1]{\texttt{#1}}
\providecommand{\urlprefix}{URL }

\bibitem{turro2009principles}
N.J. Turro, V. Ramamurthy, V. Ramamurthy and J.C. Scaiano, \emph{Principles of
  molecular photochemistry: an introduction}   (2009).

\bibitem{Scholes2011lessons}
G.D. Scholes, G.R. Fleming, A. Olaya-Castro and R. {Van Grondelle},  Nature
  Chemistry  \textbf{3} (10), 763--774 (2011).

\bibitem{Bredas2016photovoltaic}
J.L. Br{\'{e}}das, E.H. Sargent and G.D. Scholes,  Nat. Mater.  \textbf{16}
  (1), 35--44 (2016).

\bibitem{welin2017photosensitized}
E.R. Welin, C. Le, D.M. Arias-Rotondo, J.K. McCusker and D.W. MacMillan,
  Science  \textbf{355} (6323), 380--385 (2017).

\bibitem{nilsson2005electronic}
A. Nilsson, L. Pettersson, B. Hammer, T. Bligaard, C.H. Christensen and J.K.
  N{\o}rskov,  Catal. Lett.  \textbf{100}, 111--114 (2005).

\bibitem{harris1989symmetry}
D.C. Harris and M.D. Bertolucci, \emph{Symmetry and spectroscopy: an
  introduction to vibrational and electronic spectroscopy}   (1989).

\bibitem{Serrano-Andres1993}
L. Serrano-Andr{\'{e}}s, M. Merch{\'{a}}n, I. Nebot-Gil, B.O. Roos and M.
  F{\"{u}}lscher,  J. Am. Chem. Soc.  \textbf{115} (14), 6184--6197 (1993).

\bibitem{Baranov1986}
V.I. Baranov, G.N. Ten and L.A. Gribov,  J. Mol. Struct. THEOCHEM  \textbf{137}
  (1-2), 91--111 (1986).

\bibitem{Dreuw2005single-ref}
A. Dreuw and M. Head-Gordon,  Chem. Rev.  \textbf{105} (11), 4009--4037 (2005).

\bibitem{Foresman1992toward}
J.B. Foresman, M. Head-Gordon, J.A. Pople and M.J. Frisch,  J. Phys. Chem.
  \textbf{96} (1), 135--149 (1992).

\bibitem{HeadGordon1994doublescorr}
M. Head-Gordon, R.J. Rico, M. Oumi and T.J. Lee,  Chem. Phys. Lett.
  \textbf{219} (1-2), 21--29 (1994).

\bibitem{andersson1990second}
K. Andersson, P.A. Malmqvist, B.O. Roos, A.J. Sadlej and K. Wolinski,  J. Phys.
  Chem.  \textbf{94} (14), 5483--5488 (1990).

\bibitem{andersson1992second}
K. Andersson, P.{\AA}. Malmqvist and B.O. Roos,  J. Chem. Phys.  \textbf{96}
  (2), 1218--1226 (1992).

\bibitem{Krylov2008eom-ccsd}
A.I. Krylov,  Ann. Rev. Phys. Chem.  \textbf{59}, 433--462 (2008).

\bibitem{Sekino1984linear}
H. Sekino and R.J. Bartlett,  Int. J. Quantum Chem.  \textbf{26} (S18),
  255--265 (1984).

\bibitem{christiansen1995response}
O. Christiansen, H. Koch and P. J\"{o}rgensen,  J. Chem. Phys.  \textbf{103}
  (17), 7429--7441 (1995).

\bibitem{head1994doubles}
M. Head-Gordon, R.J. Rico, M. Oumi and T.J. Lee,  Chem. Phys. Lett.
  \textbf{219} (1-2), 21--29 (1994).

\bibitem{georgescu2014quantum}
I.M. Georgescu, S. Ashhab and F. Nori,  Rev. Mod. Phys.  \textbf{86} (1), 153
  (2014).

\bibitem{cao2019quantum}
Y. Cao, J. Romero, J.P. Olson, M. Degroote, P.D. Johnson, M. Kieferov{\'a},
  I.D. Kivlichan, T. Menke, B. Peropadre, N.P. Sawaya {\em{et~al.}},  Chem.
  Rev.  \textbf{119} (19), 10856--10915 (2019).

\bibitem{bauer2020quantum}
B. Bauer, S. Bravyi, M. Motta and G. Kin-Lic~Chan,  Chem. Rev.  \textbf{120}
  (22), 12685--12717 (2020).

\bibitem{mcardle2020quantum}
S. McArdle, S. Endo, A. Aspuru-Guzik, S.C. Benjamin and X. Yuan,  Rev. Mod.
  Phys.  \textbf{92}, 015003 (2020).

\bibitem{motta2021emerging}
M. Motta and J.E. Rice,  WIREs Comput. Mol. Sci  p. e1580 (2021).

\bibitem{kandala2017hardware}
A. Kandala, A. Mezzacapo, K. Temme, M. Takita, M. Brink, J.M. Chow and J.M.
  Gambetta,  Nature  \textbf{549} (7671), 242--246 (2017).

\bibitem{higgott2019variational}
O. Higgott, D. Wang and S. Brierley,  Quantum  \textbf{3}, 156 (2019).

\bibitem{gao2021applications}
Q. Gao, G.O. Jones, M. Motta, M. Sugawara, H.C. Watanabe, T. Kobayashi, E.
  Watanabe, Y.y. Ohnishi, H. Nakamura and N. Yamamoto,  npj Comput. Mater.
  \textbf{7} (1), 1--9 (2021).

\bibitem{mcclean2017subspace}
J.R. McClean, M.E. Kimchi-Schwartz, J. Carter and W.A. de~Jong,  Phys. Rev. A
  \textbf{95}, 042308 (2017).

\bibitem{colless2018computation}
J.I. Colless, V.V. Ramasesh, D. Dahlen, M.S. Blok, M.E. Kimchi-Schwartz, J.R.
  McClean, J. Carter, W.A. de~Jong and I. Siddiqi,  Phys. Rev. X  \textbf{8}
  (1), 011021 (2018).

\bibitem{smart2021quantum}
S.E. Smart and D.A. Mazziotti,  Phys. Rev. Lett.  \textbf{126} (7), 070504
  (2021).

\bibitem{ollitrault2020quantum}
P.J. Ollitrault, A. Kandala, C.F. Chen, P.K. Barkoutsos, A. Mezzacapo, M.
  Pistoia, S. Sheldon, S. Woerner, J.M. Gambetta and I. Tavernelli,  Phys. Rev.
  Research  \textbf{2} (4), 043140 (2020).

\bibitem{huang2022variational}
K. Huang, X. Cai, H. Li, Z.Y. Ge, R. Hou, H. Li, T. Liu, Y. Shi, C. Chen, D.
  Zheng {\em{et~al.}},  J. Phys. Chem. Lett.  \textbf{13}, 9114--9121 (2022).

\bibitem{motta2023quantum}
M. Motta, G.O. Jones, J.E. Rice, T.P. Gujarati, R. Sakuma, I. Liepuoniute, J.M.
  Garcia and Y.y. Ohnishi,  Chem. Sci.  \textbf{14} (11), 2915--2927 (2023).

\bibitem{eddins2022doubling}
A. Eddins, M. Motta, T.P. Gujarati, S. Bravyi, A. Mezzacapo, C. Hadfield and S.
  Sheldon,  PRX Quantum  \textbf{3} (1), 010309 (2022).

\bibitem{somma2002simulating}
R. Somma, G. Ortiz, J.E. Gubernatis, E. Knill and R. Laflamme,  Phys. Rev. A
  \textbf{65} (4), 042323 (2002).

\bibitem{tammaro2022n}
A. Tammaro, D.E. Galli, J.E. Rice and M. Motta,  J. Phys. Chem. A
  \textbf{127}, 817--827 (2023).

\bibitem{sun2018pyscf}
Q. Sun, T.C. Berkelbach, N.S. Blunt, G.H. Booth, S. Guo, Z. Li, J. Liu, J.D.
  McClain, E.R. Sayfutyarova, S. Sharma {\em{et~al.}},  WIREs Comput. Mol. Sci
  \textbf{8} (1), e1340 (2018).

\bibitem{sun2020recent}
Q. Sun {\em{et~al.}},  J. Chem. Phys.  \textbf{153} (2), 024109 (2020).

\bibitem{shao2015advances}
Y. Shao, Z. Gan, E. Epifanovsky, A.T. Gilbert, M. Wormit, J. Kussmann, A.W.
  Lange, A. Behn, J. Deng, X. Feng {\em{et~al.}},  Mol. Phys.  \textbf{113}
  (2), 184--215 (2015).

\bibitem{entanglement-forging}
L. Bello, A.M. Bra\'{n}czyk, S. Bravyi, A. Eddins, J. Gacon, T.P. Gujarati, I.
  Hamamura, T. Imamichi, C. Johnson, I. Liepuoniute, M. Motta, M. Rossmannek,
  T.L. Scholten, I. Sitdikov and S. Woerner,  GitHub  \textbf{0.1.0} (2021),
  \url{https://github.com/qiskit-community/prototype-entanglement-forging}.

\bibitem{aleksandrowicz2019qiskit}
G. Aleksandrowicz, T. Alexander, P. Barkoutsos, L. Bello, Y. Ben-Haim, D.
  Bucher, F. Cabrera-Hern{\'a}ndez, J. Carballo-Franquis, A. Chen, C. Chen
  {\em{et~al.}},  Zenodo  \textbf{16} (2019),
  \url{https://zenodo.org/record/2562111#.XhA8qi2ZPyI}.

\bibitem{nation2021scalable}
P.D. Nation, H. Kang, N. Sundaresan and J.M. Gambetta,  PRX Quantum  \textbf{2}
  (4), 040326 (2021).

\bibitem{viola1998dynamical}
L. Viola and S. Lloyd,  Phys. Rev. A  \textbf{58} (4), 2733 (1998).

\bibitem{kofman2001universal}
A. Kofman and G. Kurizki,  Phys. Rev. Lett.  \textbf{87} (27), 270405 (2001).

\bibitem{biercuk2009optimized}
M.J. Biercuk, H. Uys, A.P. VanDevender, N. Shiga, W.M. Itano and J.J.
  Bollinger,  Nature  \textbf{458} (7241), 996--1000 (2009).

\bibitem{rost2020simulation}
B. Rost, B. Jones, M. Vyushkova, A. Ali, C. Cullip, A. Vyushkov and J.
  Nabrzyski,  arXiv:2001.00794   (2020).

\bibitem{niu2022effects}
S. Niu and A. Todri-Sanial,  IEEE Trans. Quantum Eng.  \textbf{3}, 1--10
  (2022).

\bibitem{niu2022analyzing}
S. Niu and A. Todri-Sanial,  arXiv:2204.14251   (2022).

\bibitem{ezzell2022dynamical}
N. Ezzell, B. Pokharel, L. Tewala, G. Quiroz and D.A. Lidar,  arXiv:2207.03670
   (2022).

\bibitem{huggins2021efficient}
W.J. Huggins, J.R. McClean, N.C. Rubin, Z. Jiang, N. Wiebe, K.B. Whaley and R.
  Babbush,  npj Quantum Inf.  \textbf{7} (1), 1--9 (2021).

\bibitem{cohn2021quantum}
J. Cohn, M. Motta and R.M. Parrish,  PRX Quantum  \textbf{2} (4), 040352
  (2021).

\bibitem{d2022accuracy}
R. D’Cunha, T.D. Crawford, M. Motta and J.E. Rice,  J. Phys. Chem. A
  \textbf{127} (15), 3437--3448 (2023).

\bibitem{ryabinkin2018qubit}
I.G. Ryabinkin, T.C. Yen, S.N. Genin and A.F. Izmaylov,  J. Chem. Theory
  Comput.  \textbf{14} (12), 6317--6326 (2018).

\bibitem{schreiber2008benchmarks}
M. Schreiber, M.R. Silva-Junior, S.P. Sauer and W. Thiel,  J. Chem. Phys.
  \textbf{128} (13), 134110 (2008).

\bibitem{flicker1976electron}
W.M. Flicker, O.A. Mosher and A. Kuppermann,  J. Chem. Phys.  \textbf{64} (4),
  1315--1321 (1976).

\bibitem{Palmer1995}
M.H. Palmer, I.C. Walker, C.C. Ballard and M.F. Guest,  Chem. Phys.
  \textbf{192} (2), 111--125 (1995).

\bibitem{van1976triplet}
E. Van~Veen,  Chem. Phys. Lett.  \textbf{41} (3), 535--539 (1976).

\bibitem{Cai2000}
Z.L. Cai and J.R. Reimers,  J. Phys. Chem. A  \textbf{104} (36), 8389--8408
  (2000).

\bibitem{Walker1990}
I.C. Walker, M.H. Palmer and A. Hopkirk,  Chem. Phys.  \textbf{141} (2-3),
  365--378 (1990).

\bibitem{linert2016electron}
I. Linert and M. Zubek,  Eur. Phys. J. D  \textbf{70}, 1--8 (2016).

\bibitem{linert2015study}
I. Linert and M. Zubek,  Chem. Phys. Lett.  \textbf{624}, 1--5 (2015).

\bibitem{palmer1990electronic}
M.H. Palmer, I.C. Walker, M.F. Guest and A. Hopkirk,  Chem. Phys.  \textbf{147}
  (1), 19--33 (1990).

\bibitem{koch1997cc3}
H. Koch, O. Christiansen, P. J\"{o}rgensen, A.M. Sanchez~de Mer{\'a}s and T.
  Helgaker,  J. Chem. Phys.  \textbf{106} (5), 1808--1818 (1997).

\bibitem{asthana2023quantum}
A. Asthana, A. Kumar, V. Abraham, H. Grimsley, Y. Zhang, L. Cincio, S. Tretiak,
  P.A. Dub, S.E. Economou, E. Barnes {\em{et~al.}},  Chem. Sci.  \textbf{14}
  (9), 2405--2418 (2023).

\bibitem{tilly2021reduced}
J. Tilly, P. Sriluckshmy, A. Patel, E. Fontana, I. Rungger, E. Grant, R.
  Anderson, J. Tennyson and G.H. Booth,  Phys. Rev. Research  \textbf{3} (3),
  033230 (2021).

\bibitem{choi2023measurement}
S. Choi and A.F. Izmaylov,  J. Chem. Theory Comput.   (2023).

\bibitem{rice2021quantum}
J.E. Rice, T.P. Gujarati, M. Motta, T.Y. Takeshita, E. Lee, J.A. Latone and
  J.M. Garcia,  J. Chem. Phys.  \textbf{154} (13), 134115 (2021).

\bibitem{o2022purification}
T.E. O'Brien, G. Anselmetti, F. Gkritsis, V. Elfving, S. Polla, W.J. Huggins,
  O. Oumarou, K. Kechedzhi, D. Abanin, R. Acharya {\em{et~al.}},
  arXiv:2210.10799   (2022).

\bibitem{google2020hartree}
F. Arute, K. Arya, R. Babbush, D. Bacon, J.C. Bardin, R. Barends, S. Boixo, M.
  Broughton, B.B. Buckley {\em{et~al.}},  Science  \textbf{369} (6507),
  1084--1089 (2020).

\bibitem{zhao2022orbital}
L. Zhao, J. Goings, K. Wright, J. Nguyen, J. Kim, S. Johri, K. Shin, W. Kyoung,
  J.I. Fuks, J.K.K. Rhee {\em{et~al.}},  arXiv:2212.02482   (2022).

\bibitem{guo2022scalable}
S. Guo, J. Sun, H. Qian, M. Gong, Y. Zhang, F. Chen, Y. Ye, Y. Wu, S. Cao, K.
  Liu {\em{et~al.}},  arXiv:2212.08006   (2022).

\bibitem{huggins2022unbiasing}
W.J. Huggins, B.A. O’Gorman, N.C. Rubin, D.R. Reichman, R. Babbush and J.
  Lee,  Nature  \textbf{603} (7901), 416--420 (2022).

\bibitem{motta2018ab}
M. Motta and S. Zhang,  WIREs Comput. Mol. Sci  \textbf{8} (5), e1364 (2018).

\end{thebibliography}

\end{document}